\documentclass[11pt]{article}
\usepackage{moriond,epsfig,multicol}

\def\be{\begin{equation}}
\def\ee{\end{equation}}
\def\bea{\begin{eqnarray}}
\def\eea{\end{eqnarray}}

\newcommand{\met}       {\mbox{$\not\!\!E_T$}}
\newcommand{\pt}        {\mbox{p$_{\mathrm T}$}}
\newcommand{\rar}       {\rightarrow}
\newcommand{\rargap}    {\mbox{ $\rightarrow$ }}
\newcommand{\ttbar}     {\mbox{$t\bar{t}$}}
\newcommand{\ppbar}     {\mbox{$p\bar{p}$}}
\def \dzero    {D\O} 

\begin{document}
\vspace*{4cm}
\title{Evidence for single top quark production at \dzero}

\author{Ar\'an Garc\'\i a-Bellido}

\address{Department of Physics, University of Washington, \\ Seattle, WA
98195, USA}

\maketitle\abstracts{
The results of the first analysis to show evidence for production of single top
quarks are presented.  Using
0.9~fb$^{-1}$ of data collected with the \dzero~detector at the
Fermilab Tevatron, the analysis is performed in the electron+jets and muon+jets
decay modes, taking 
special care in modeling the large backgrounds, applying a new powerful
$b$-quark tagging algorithm and using three multivariate techniques to extract the small signal
in the data.  
The combined measured production cross section is $4.8 \pm 1.3$~pb.  The probability to
measure a cross section at this value or higher in the absence of a signal is
$0.027\%$, corresponding to a 3.5~standard deviation significance. 
}

\section{Introduction}
The top quark is very special fermion: it is by far the heaviest
elementary particle discovered so far. Its large mass gives it a prominent role in the
mechanism of electroweak symmetry breaking, since the Higgs boson coupling to
fermions is proportional to their mass, and hence top quarks and Higgs bosons
have a coupling strength of order unity. Top quarks also decay before they
hadronize and thus pass their kinematic properties to the decay products, that can be
then used to study the nature of ``bare'' quarks.
The study of this elusive quark has provided us with some new insights into its
properties recently, like its charge of 2/3$e$, its dominant decay to a $W$
boson and a bottom quark, and its production cross section and mass, these last
two being measured with ever increasing 
precision since its discovery.

Top quarks were first observed produced in {\ttbar} pairs via the strong interaction at
the Tevatron collider in 1995. In the SM, top quarks are
also expected to be produced via the exchange of a $W$~boson in $s$- or
$t$-channel. The final state in these channels
thus consists of one ``single'' top quark together with a $b$ quark in the
$s$-channel ($tb$) and an additional light quark in the $t$-channel
($tqb$). Single top quarks can also be produced in association with a $W$~boson
($tW$), but the cross section for this process at the Tevatron is very small and
will be ignored here.  The
next-to-leading order prediction for the $s$-channel single top quark
cross section is $\sigma({\ppbar}{\rargap}tb+X) = 0.88\pm0.11$~pb, and
for the $t$-channel process, the prediction is $\sigma({\ppbar}
{\rargap}tqb+X) = 1.98\pm0.25$~pb~\cite{singletop-xsec-sullivan}.

Both the CDF and \dzero~collaborations have performed searches for this
process in the past~\cite{cdf}. In Run II, the best published $95\%$~C.L. upper limits are
$\sigma({\ppbar}{\rargap}tb+X)<6.4$~pb and
$\sigma({\ppbar}{\rargap}tqb+X)<5.0$~pb~\cite{limits-d0}.   
The analysis presented here, described in more detail in
Ref.~\cite{Abazov:2006gd},
 draws many techniques and experience from the previous
\dzero~analyses, where it was made clear that multivariate techniques are
necessary to be sensitive to the SM production cross section with limited
data statistics.

\section{Event selection}
The data used in this search was collected from 2002 to 2005 with triggers that
required an electron or a muon and at least one jet. The average trigger
efficiency for e+jets events is 86\% and 87\% in the $t$- and $s$-channels, and
for $\mu$+jets, 82\% and 87\% in $t$- and $s$-channels signal samples respectively. 
Events are required to have exactly one isolated electron (muon) with
$p_T>15$~GeV (18~GeV) within $|\eta|<1.1$ (2.0), and $\met > 15$~GeV. 
Events are also required to contain two, three or four jets, using a cone
algorithm with radius $\mathcal{R} =
\sqrt{(\Delta y)^2+(\Delta\phi)^2} = 0.5$ (where $y$ is rapidity and
$\phi$ is azimuthal angle) to cluster energy deposits in the
calorimeter. The leading jet has $p_T>25$~GeV and $|\eta|<2.5$, the
second leading jet has $p_T>20$~GeV and $|\eta|<3.4$, and subsequent
jets have $p_T>15$~GeV and $|\eta|<3.4$.

The selection requirements described above achieve a summed signal to
background ratio of around 1:180, with an acceptance of around 5.0\% and 4.5\% for
$s$- and $t$-channel events respectively. 
To enhance the signal content of the selection, at least one jet in the event is
required to be identified as a $b$-quark jet. 
A neural network~\cite{nntag} has been used to identify jets originating from long-lived
$b$~hadrons. The variables used to identify such jets rely on the presence and
characteristics of a secondary vertex and tracks with high impact
parameters inside the jet. These variables are, ranked in order of
separation power: (i) decay length significance of the Secondary Vertex Tagger (SVT); (ii)
weighted combination of the tracks' IP significances; (iii) JLIP
probability (that the jet originates from the primary vertex based on the Jet
LIfetime Probability algorithm); (iv) $\chi^2$ per
degree of freedom of the SVT secondary vertex; (v) number of tracks
used to reconstruct the secondary vertex; (vi) mass of the secondary
vertex; and (vii) number of secondary vertices found inside the jet.
For a $0.5\%$ light-jet (mis)tag rate, we
obtain a $50\%$ average tag rate in data for $b$~jets with
$|\eta|<2.4$. 

\begin{figure}[htbp]
\begin{center}
\includegraphics[width=0.32\textwidth]{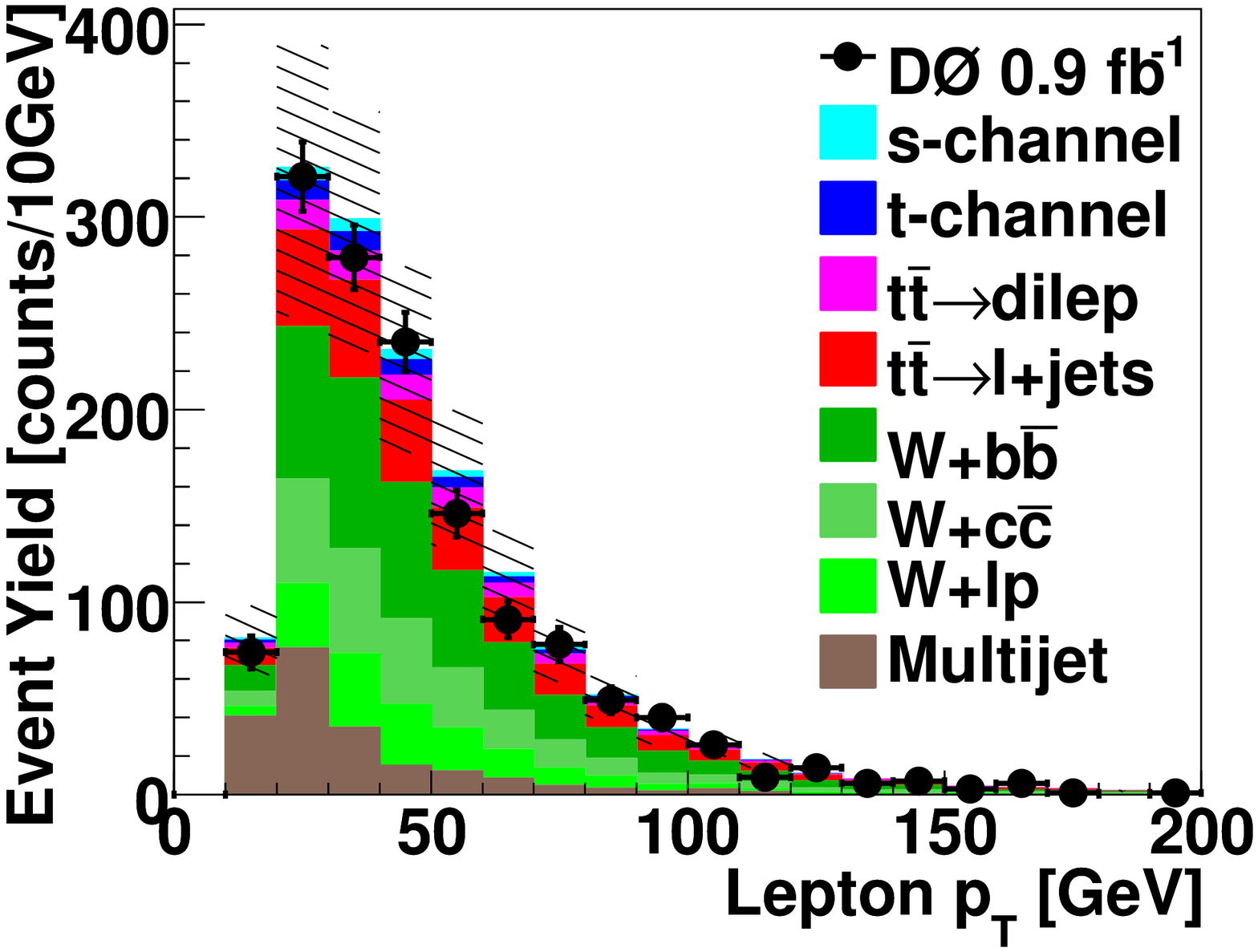}
\includegraphics[width=0.32\textwidth]{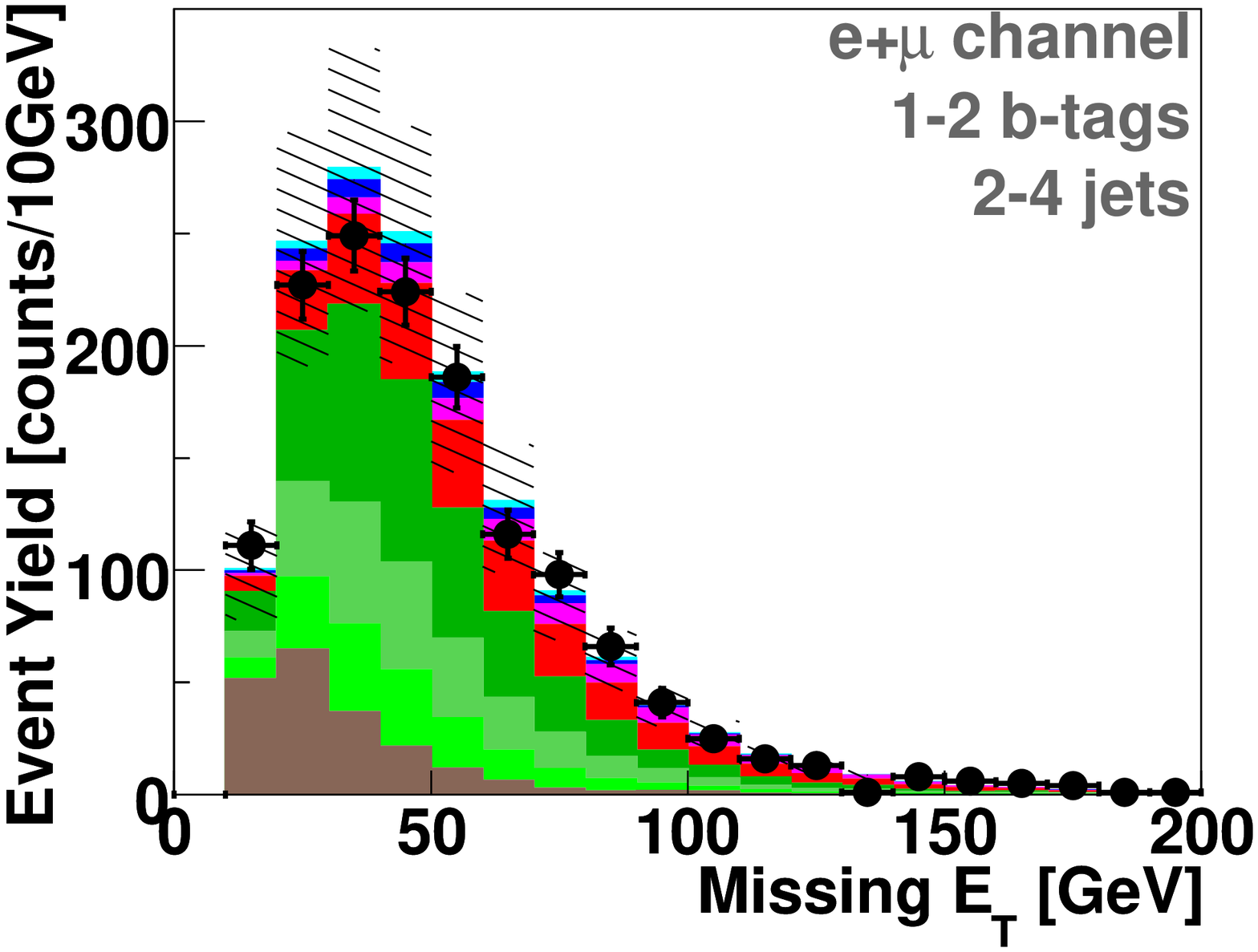}
\includegraphics[width=0.32\textwidth]{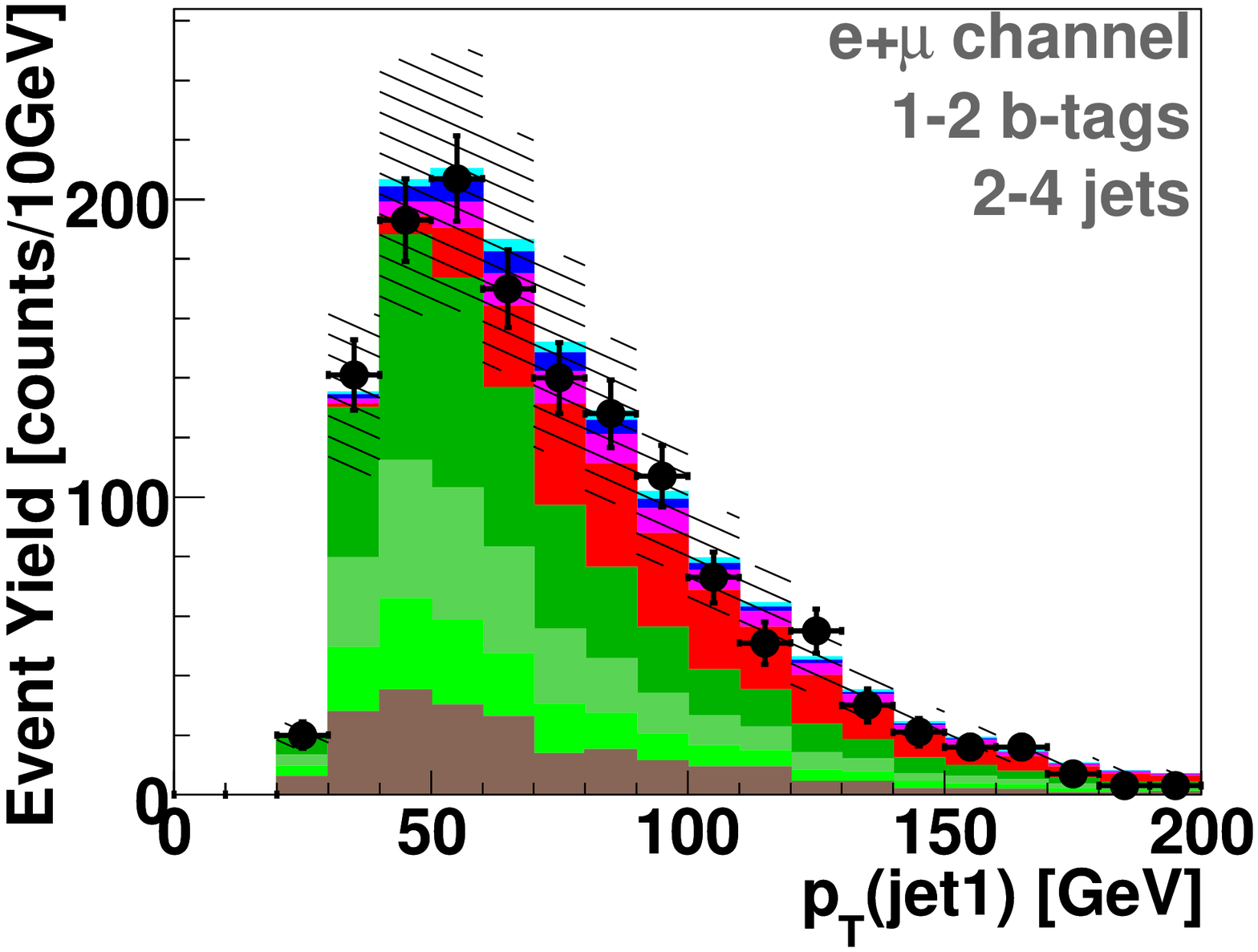}
\includegraphics[width=0.32\textwidth]{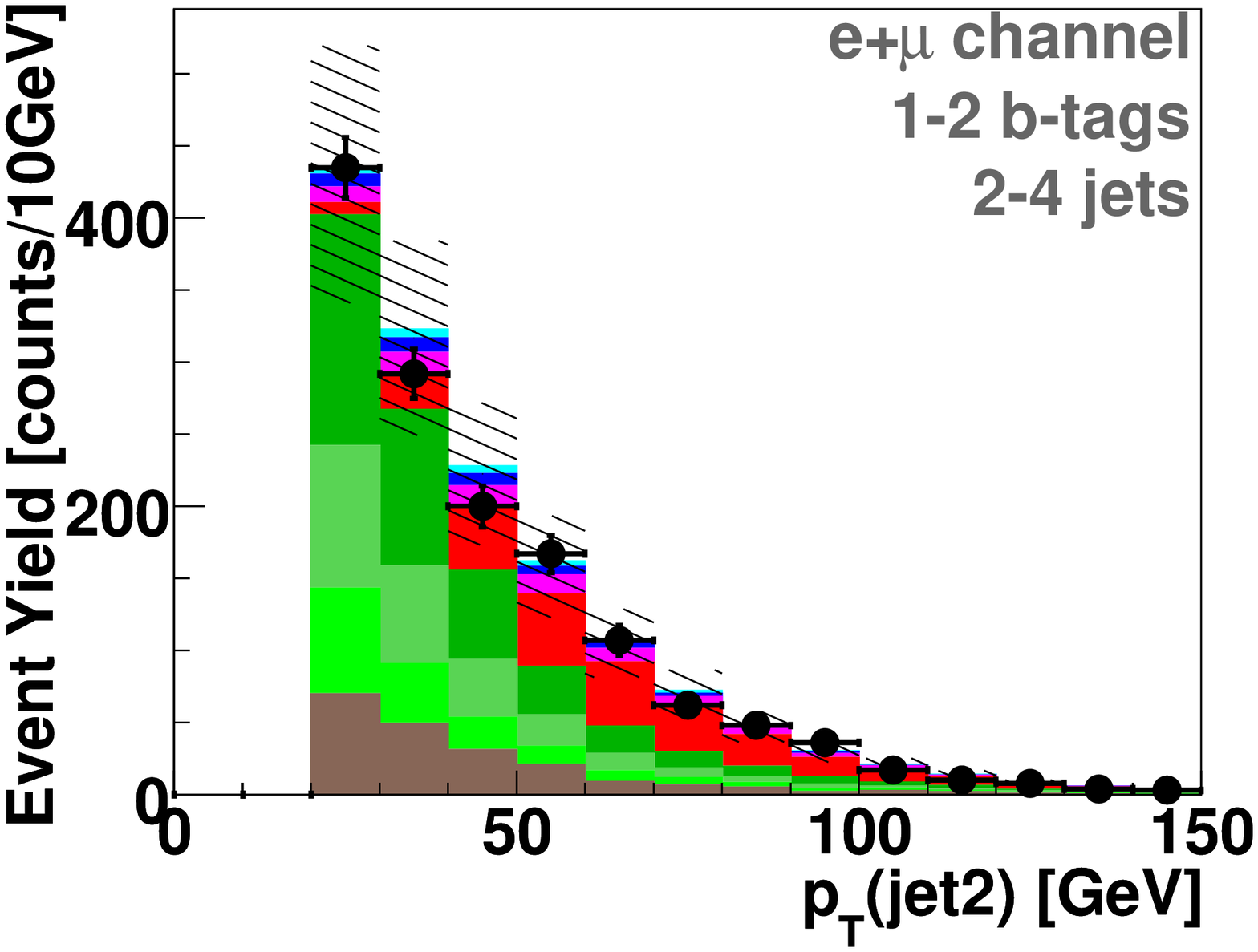}
\includegraphics[width=0.32\textwidth]{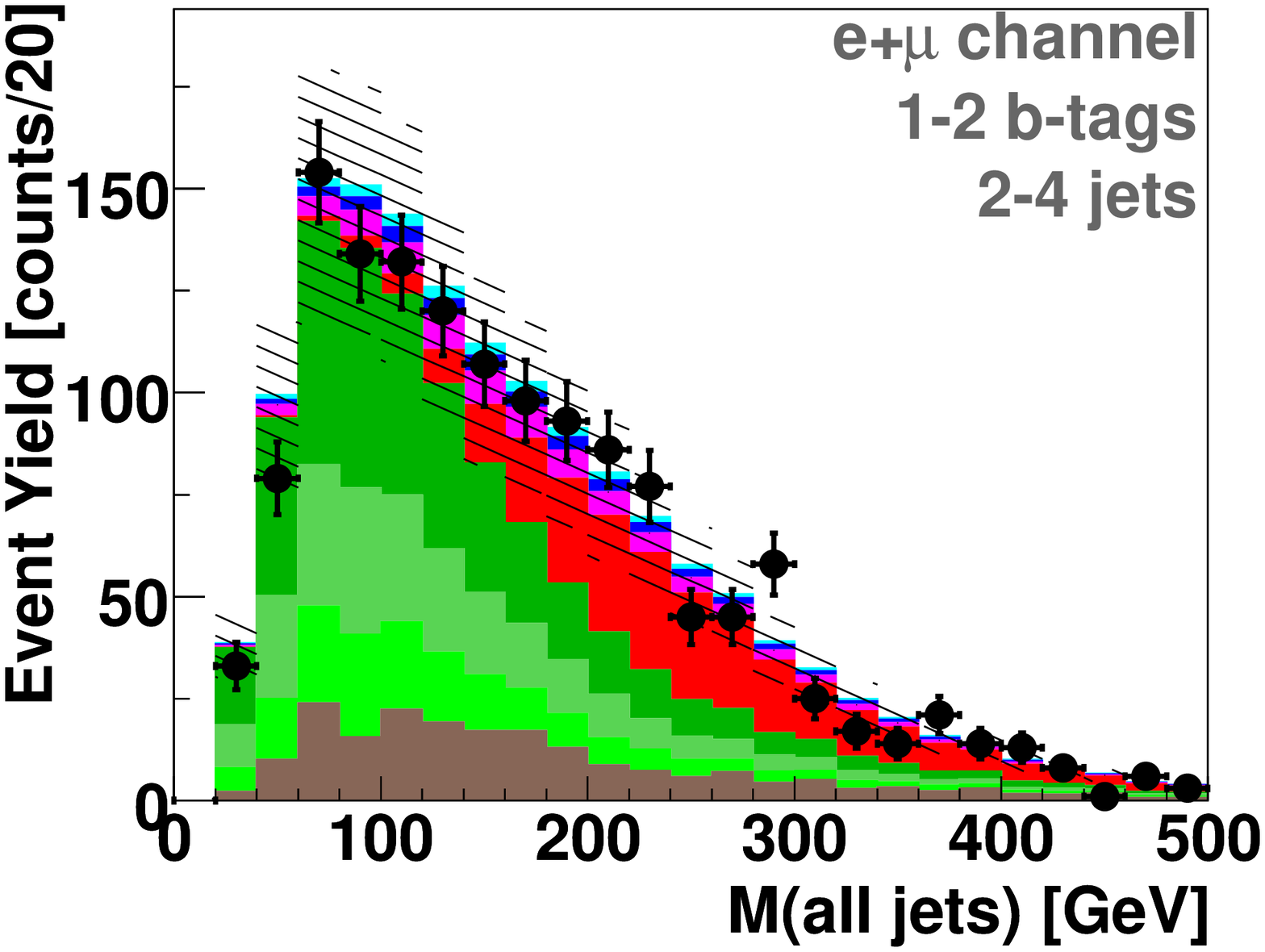}
\includegraphics[width=0.32\textwidth]{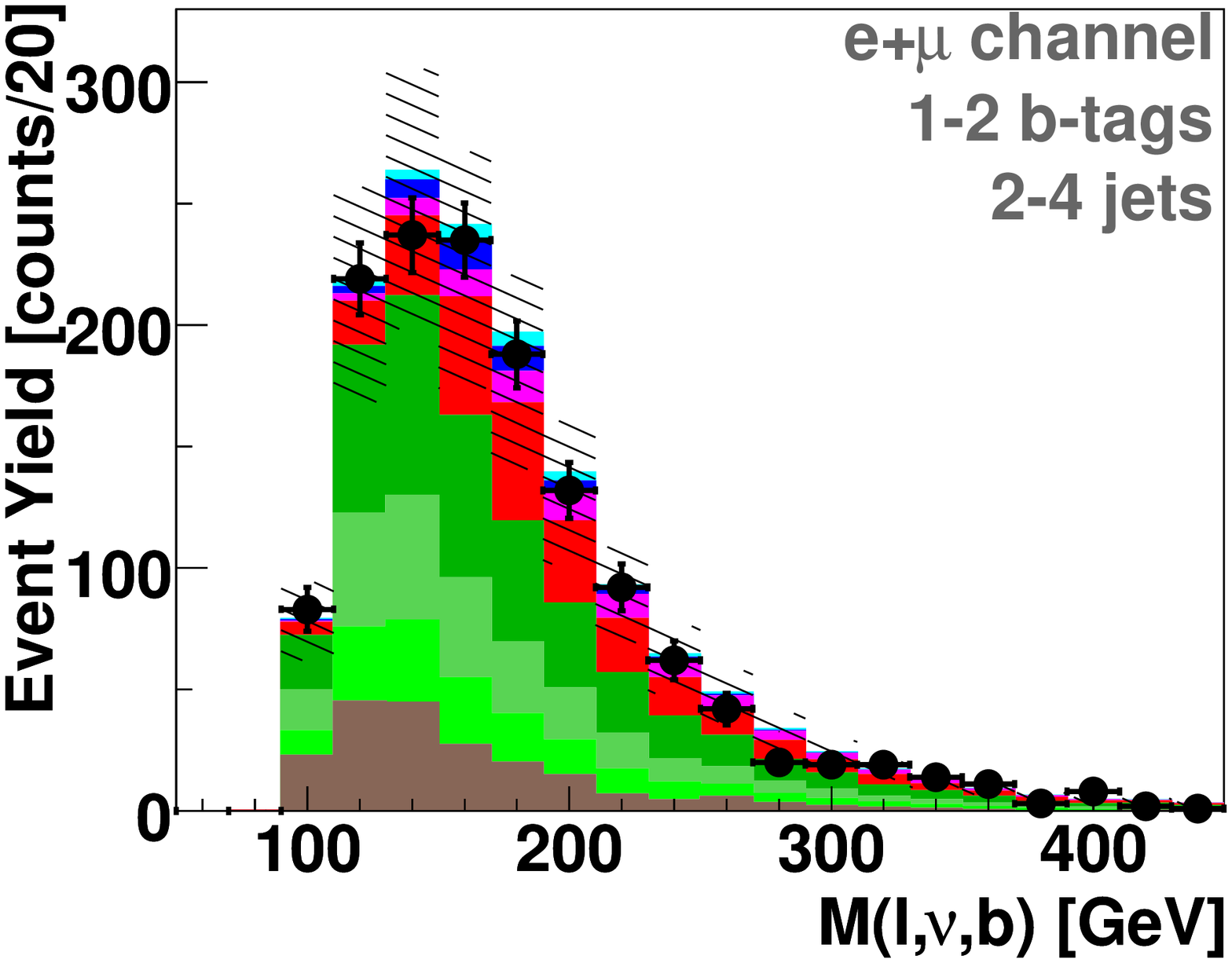}
\end{center}
\vspace{-0.1in}
\caption[dtplots]{Data-background agreement distributions for all channels
combined for the lepton \pt, \met, leading jet \pt, second leading jet \pt, the invariant mass of
all jets, and the invariant mass of the reconstructed $W$ together with the
$b$-tagged jet. The hashed bands show the $\pm 1$ standard deviation uncertainty on the background.}
\label{sel_plots}
\end{figure}

By requiring events to have at least one $b$-tagged jet, the signal to
background ratio is enhanced to 1:22 on the sum of all channels, and the most
sensitive channel, with two jets and one $b$-tagged jet, reaches a signal to
background ratio of 1:10. The acceptance after $b$-tagging is reduced to 3.2\%
in the $s$-channel and 2.1\% in the $t$-channel. The final event yields after
$b$-tagging are given in Table~\ref{event-yields}, shown separated only by jet multiplicity.  
Figure~\ref{sel_plots} shows
the data-background agreement in six basic distributions for all the channels
combined. 
\begin{table}[htbp]
\caption[eventyields]{Numbers of expected and observed events in
0.9~fb$^{-1}$ for $e$ and $\mu$, 1 $b$~tag and 2 $b$~tag channels
combined. The total background uncertainties are smaller than the
component uncertainties added in quadrature because of anticorrelation
between the $W$+jets and multijet backgrounds resulting from the
background normalization procedure.}
\label{event-yields}
\begin{center}
\begin{tabular}{l@{\hspace{0.5cm}}r@{~~$\pm$~~}l@{}%
                 @{\hspace{0.5cm}}r@{~~$\pm$~~}l@{}%
                 @{\hspace{0.5cm}}r@{~~$\pm$~~}l@{}}
Source           & \multicolumn{2}{c}{2 jets}
                 & \multicolumn{2}{c}{3 jets}
                 & \multicolumn{2}{c}{4 jets} \\
\hline\hline	     
$tb$                      &  16  &   3  &   8  &  2  &   2  &  1  \\
$tqb$                     &  20  &   4  &  12  &  3  &   4  &  1  \\
\hline                                                              
${\ttbar}{\rar}\ell\ell$  &  39  &   9  &  32  &  7  &  11  &  3  \\
${\ttbar}{\rar}\ell$+jets &  20  &   5  & 103  & 25  & 143  & 33  \\
$Wb\bar{b}$               & 261  &  55  & 120  & 24  &  35  &  7  \\
$Wc\bar{c}$               & 151  &  31  &  85  & 17  &  23  &  5  \\
$Wjj$                     & 119  &  25  &  43  &  9  &  12  &  2  \\
Multijets                 &  95  &  19  &  77  & 15  &  29  &  6  \\
\hline                                                              
Total background          & 686  &  41  & 460  & 39  & 253  & 38  \\
Data             & \multicolumn{2}{c}{697}
                 & \multicolumn{2}{c}{455}
                 & \multicolumn{2}{c}{246}
\end{tabular}
\end{center}
\end{table}

\section{Background normalization and agreement with data} 
Looking at Tab.~\ref{event-yields}, it is evident that if we are to reliably
extract such a small signal from such large backgrounds (around 62 expected signal events
from a total of 1,398 data events,  where the error on the total background prediction
is larger than the expected signals) we need to make sure the backgrounds are
properly modeled in both their kinematics and overall normalization. 

The {\ttbar} background is normalized to the integrated luminosity times the
predicted {\ttbar} cross section of $6.8^{+0.6}_{-0.5}$~pb~\cite{ttbar-xsec-1}.
The $W$+jets background, together with the multijet background, is normalized to
the data in each channel (defined by lepton flavor and jet multiplicity) before
$b$-tagging. 

A second normalization to data is performed in order to determine how much of
the total $W$+jets sample is actually made of heavy-flavor jets, i.e. the
fraction of $Wb\bar{b}$ and $Wc\bar{c}$ in the total $W$+jets
sample. $Wc$+jets is included in our model together with other
$W$+light-quark jets, i.e.: inside $Wjj$ in Tab.~\ref{event-yields}.
The heavy-flavor fraction is determined in our selected data on those events
in which the neural network $b$-tagger fails to find a $b$ jet of the required
quality. This sample of events is not used for the signal measurement and has
approximately the same heavy-flavor composition as the signal sample. 
We find that a constant scale factor of $1.5\pm0.45$ applied to the $Wb\bar{b}$ and
$Wc\bar{c}$ components is necessary to achieve a good description of the
data.  The uncertainty assigned to this factor covers the
expected dependence on event kinematics and the assumption that the
scale factor is the same for $Wb\bar{b}$ and $Wc\bar{c}$. 

This constant heavy-flavor scale factor which absorbs higher order
effects~\cite{wbb-wbj-nlo}, was found to give overall a very good description of
the data. It was checked that the most sensitive variables had well
described shapes, specifically those distributions expected to suffer the
largest shape dependence from higher order corrections, like the invariant mass
of the two leading jets and the $\pt$ of the $b$-tagged jet. Two control
samples, one enriched in $W$+jets events and the other in $\ttbar$ events, were
also used to check for shape disagreements in the background model description
of the data, and overall good agreement was found.

\section{Systematic errors}
The dominant contributions to the uncertainties on the backgrounds
come from: normalization of the {\ttbar} background (18\%), which
includes a term to account for the top quark mass uncertainty;
normalization of the $W$+jets and multijet backgrounds to data
(17--27\%), which includes the uncertainty on the heavy-flavor
fraction of the model; and the $b$-tagging probabilities (12--17\% for
double-tagged events). The uncertainty on the integrated luminosity is
6\%; all other sources contribute at the few percent level. The
uncertainties from the jet energy scale corrections and the
$b$-tagging probabilities affect both the shape and normalization of
the simulated distributions. 

The 30\% error assigned to the heavy-flavor fraction is by far the dominant
uncertainty in the final cross section measurement. Its impact is nevertheless
not directly 30\% because it only affects two sources of the background (before
$b$-tagging is applied) and it gets further reduced when taking into account the
first normalization of $W$+jets and QCD to data, such that in the end, the
total uncertainty on the sum of $W$+jets and QCD yields is between 17 and
27\% (depending on the channel), including the heavy-flavor factor uncertainty. 

\section{Final separation methods and cross section measurements}
\begin{figure}[htbp]
\includegraphics[width=0.32\textwidth,height=0.20\textheight]{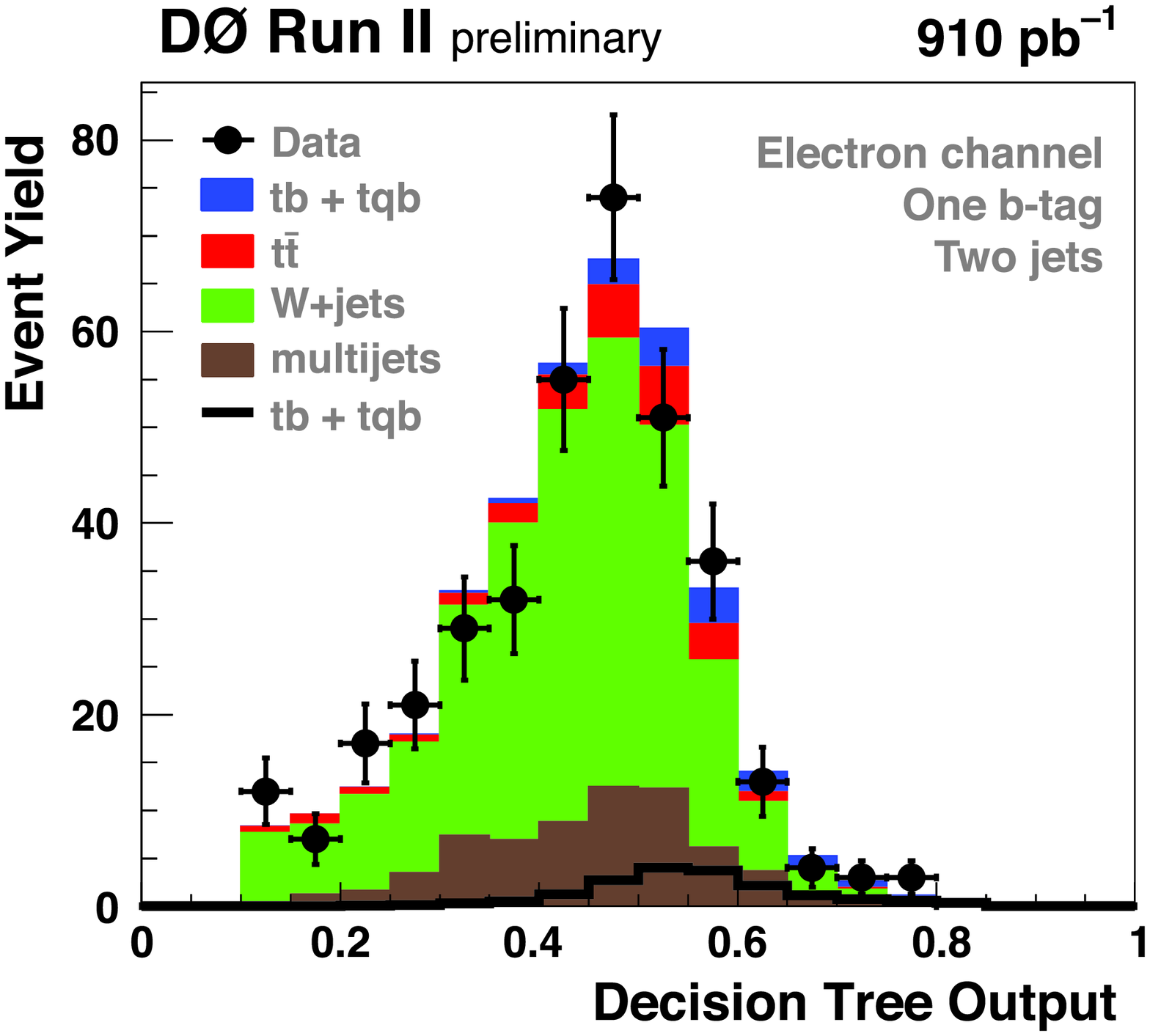}
~
\includegraphics[width=0.32\textwidth,height=0.20\textheight]{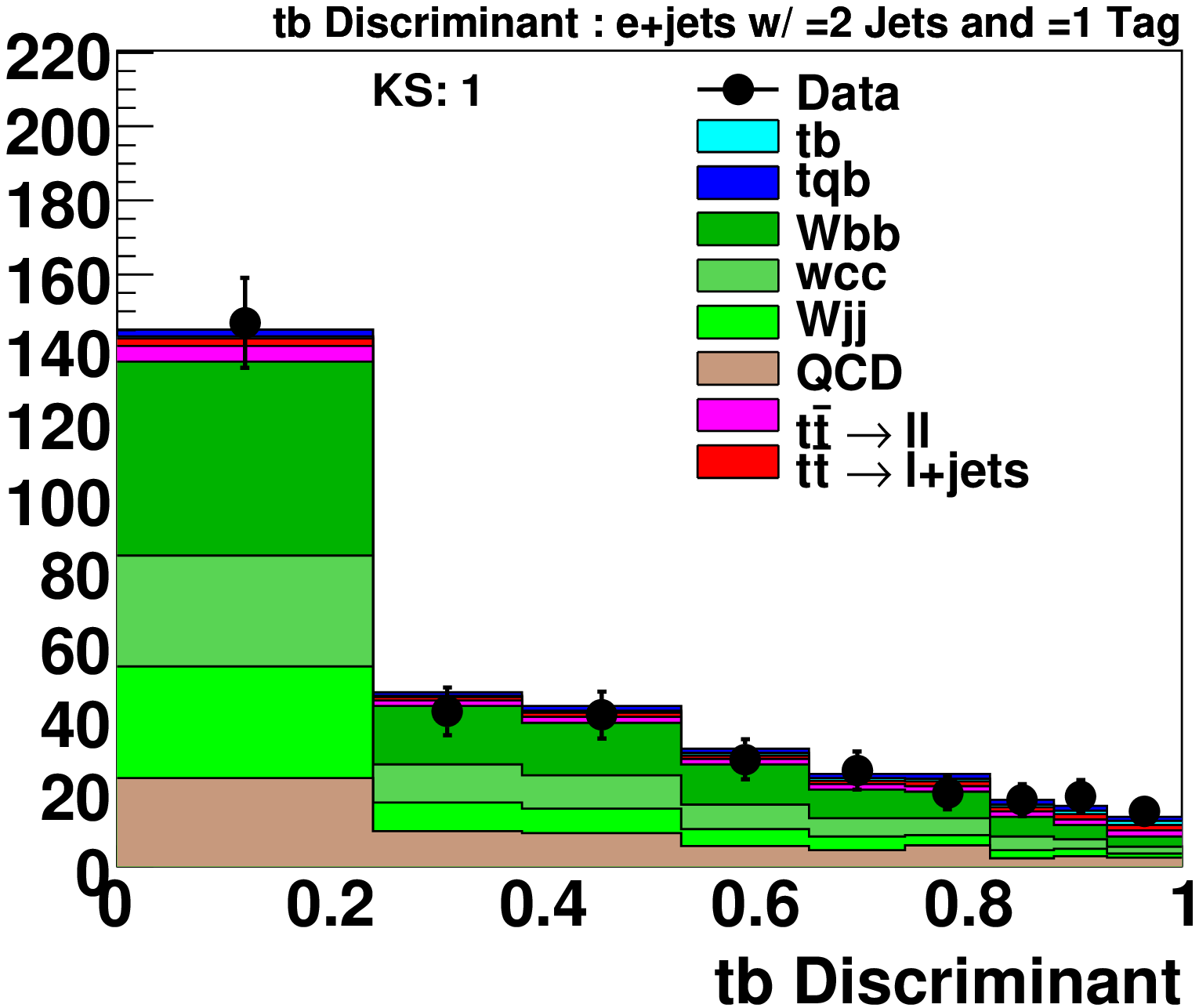}
~
\includegraphics[width=0.32\textwidth,height=0.20\textheight]{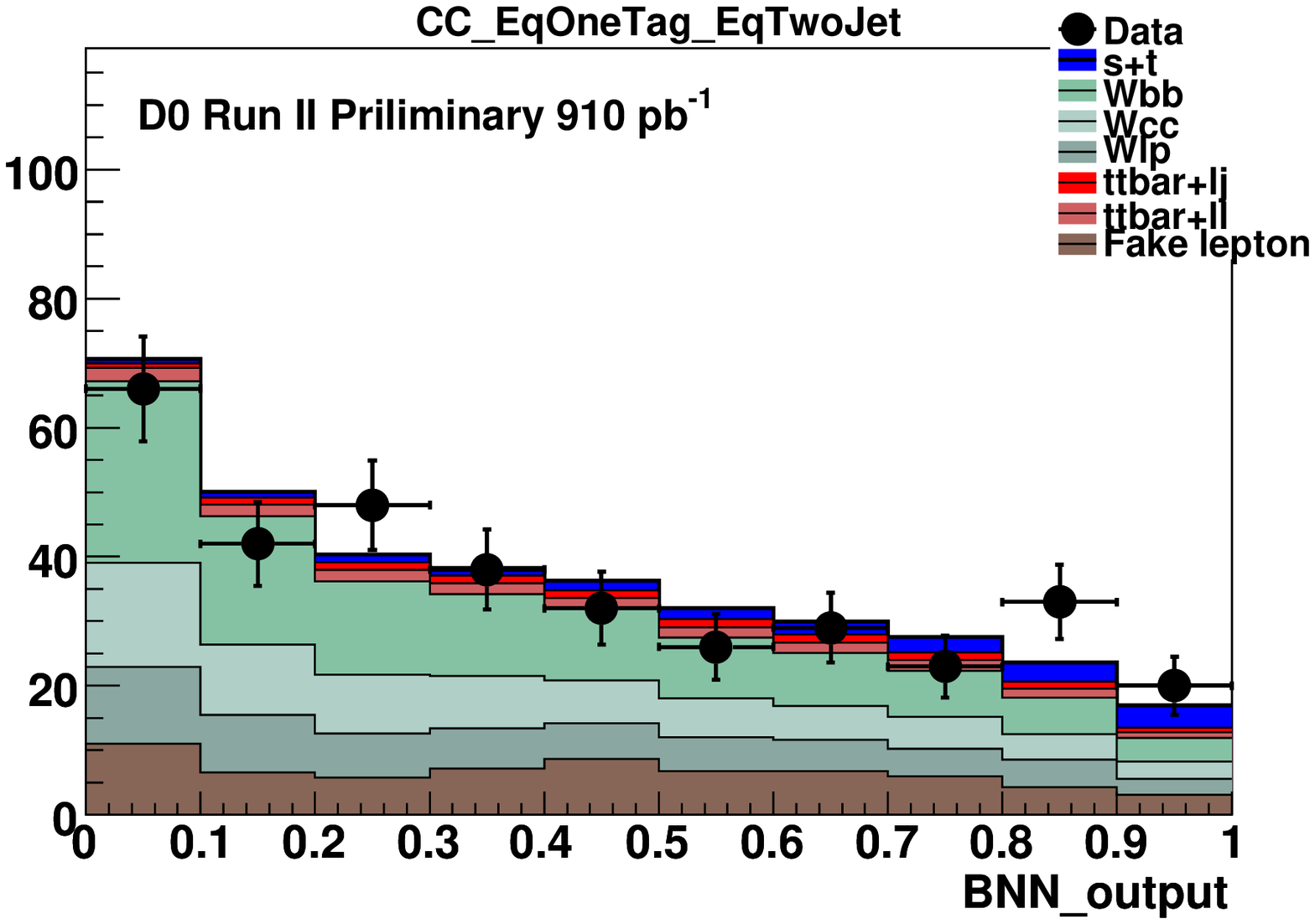}
\vspace{-0.1in}
\caption[1dout]{Output distributions in the electron channel with two jets, one
of them $b$-tagged, for the Decision Tree discriminant (left), for the Matrix Element
$tq$ discriminant (center), and for the Bayesian Neural Net discriminant
(right). }
\label{1d-out}
\end{figure}

Three
separate multivariate techniques have been employed in \dzero~to extract
the signal from the data: Boosted Decision Trees~\cite{webpage}, Matrix
Elements discriminants~\cite{5392} and Bayesian Neural Networks~\cite{5397}.
The output distributions in the most sensitive channel, shown in
Fig.~\ref{1d-out} for the three methods, also demonstrate good overall agreement between data
and background.

\begin{figure}[htbp]
\begin{minipage}{\textwidth}
 \setlength{\columnsep}{-20pt} 
\begin{multicols}{2}
\includegraphics[width=0.48\textwidth]{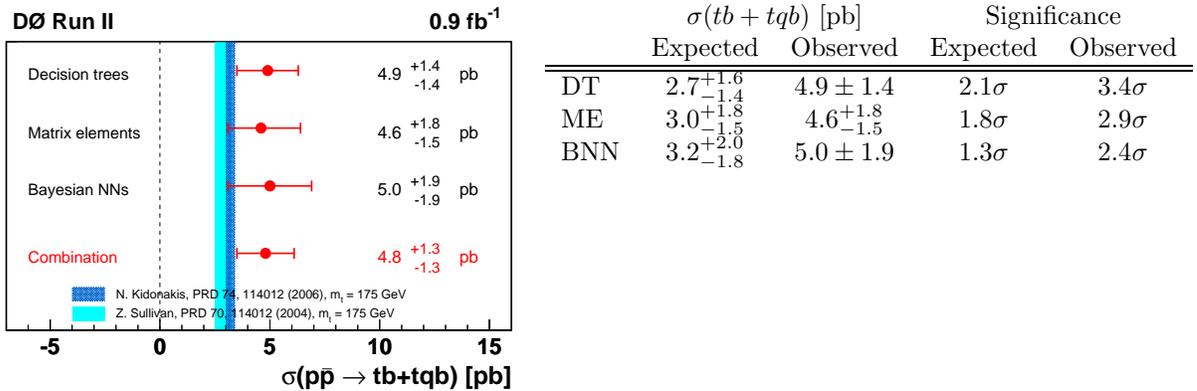}
\small\begin{center}
\vspace{1.0cm}
\begin{tabular}{lcccc}
& \multicolumn{2}{c}{$\sigma(tb+tqb)$ [pb]} & \multicolumn{2}{c}{Significance} \\
& Expected & Observed & Expected & Observed \\ \hline \hline
DT & $2.7^{+1.6}_{-1.4}$ & $4.9 \pm 1.4$       & $2.1\sigma$ & $3.4\sigma$ \\
ME & $3.0^{+1.8}_{-1.5}$ & $4.6^{+1.8}_{-1.5}$ & $1.8\sigma$ & $2.9\sigma$ \\
BNN & $3.2^{+2.0}_{-1.8}$ & $5.0 \pm 1.9$      & $1.3\sigma$ & $2.4\sigma$ \\
\end{tabular}
\end{center}
\end{multicols}
\end{minipage}
\vspace{-0.2in}
\caption[comb]{The observed results
including the combination (left); and the expected and observed cross sections and significance for
the three different multivariate analyses (right).}
\label{comb}
\end{figure}

In order to extract the maximum information from the discriminant outputs, instead of
cutting on the outputs and counting events, the full
distributions are fed into a Bayesian statistical analysis
to measure the single top quark production cross section. 
The expected and observed cross section results are summarized in Fig~\ref{comb}. 
The uncertainties include statistical and systematic components combined. 
The data statistics contribute 1.2~pb to the total 1.4~pb uncertainty on the $tb$+$tqb$
cross section for the DT analysis. The significance is measured re-running the
analysis on 70,000 pseudo-datasets generated with all the
uncertainties on the background model taken into account, but including only
background sources. Thus we obtain the probability for the background-only
hypothesis to fluctuate up to give the measured (or SM) value of the $tb$+$tqb$
cross section or greater.  

\begin{figure*}[!h!tbp]
\includegraphics[width=0.32\textwidth]{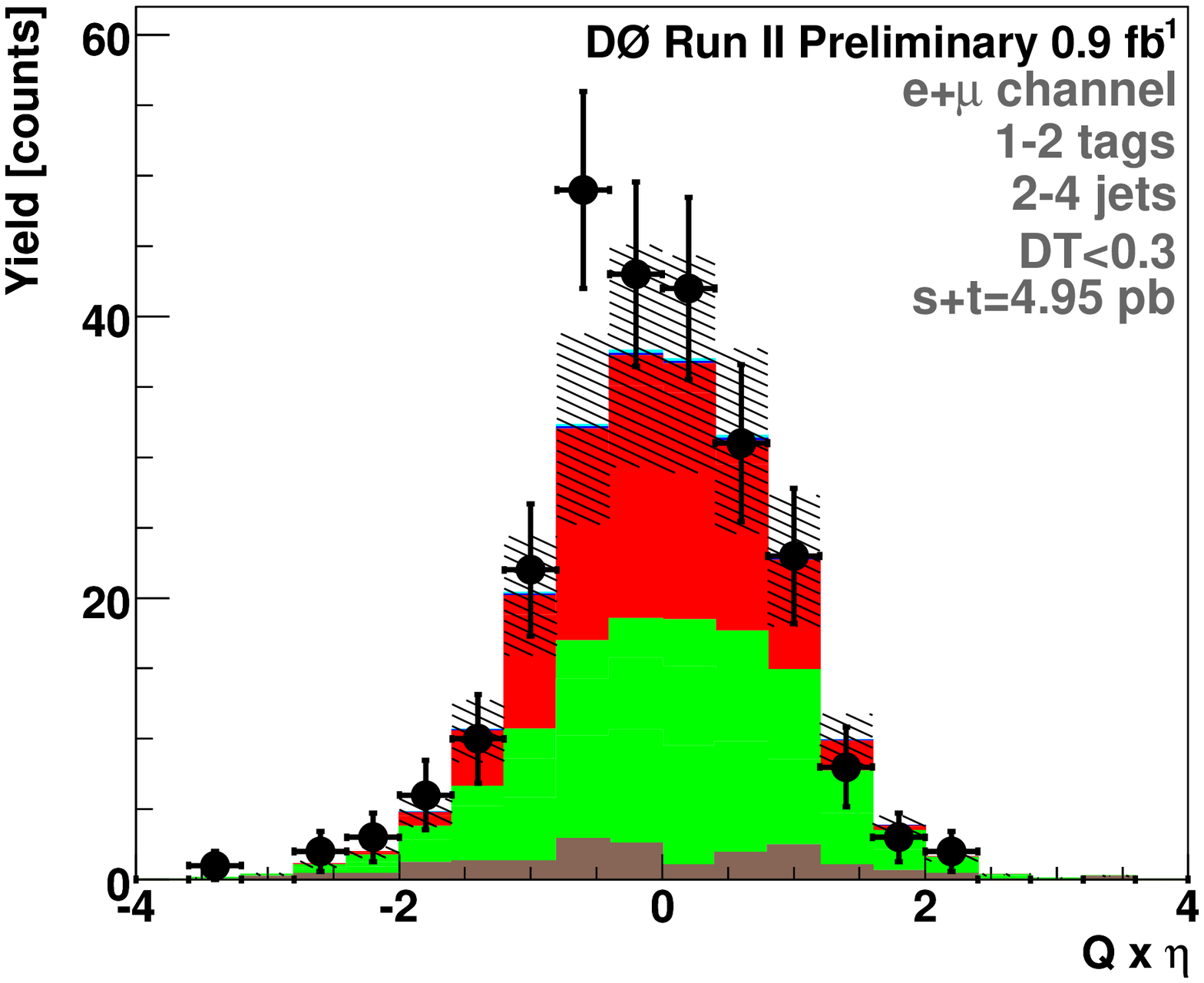}
\includegraphics[width=0.32\textwidth]{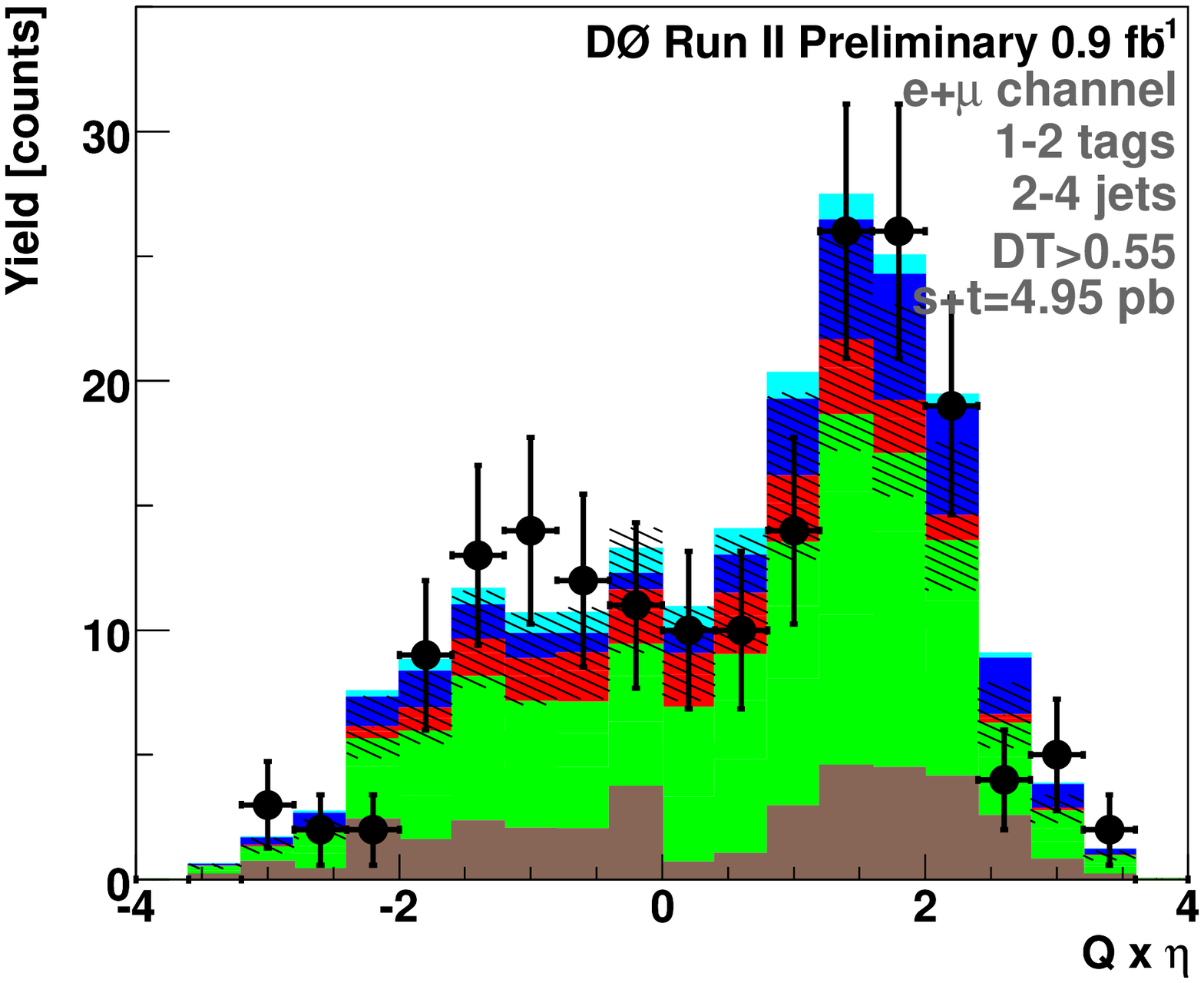}
\includegraphics[width=0.32\textwidth]{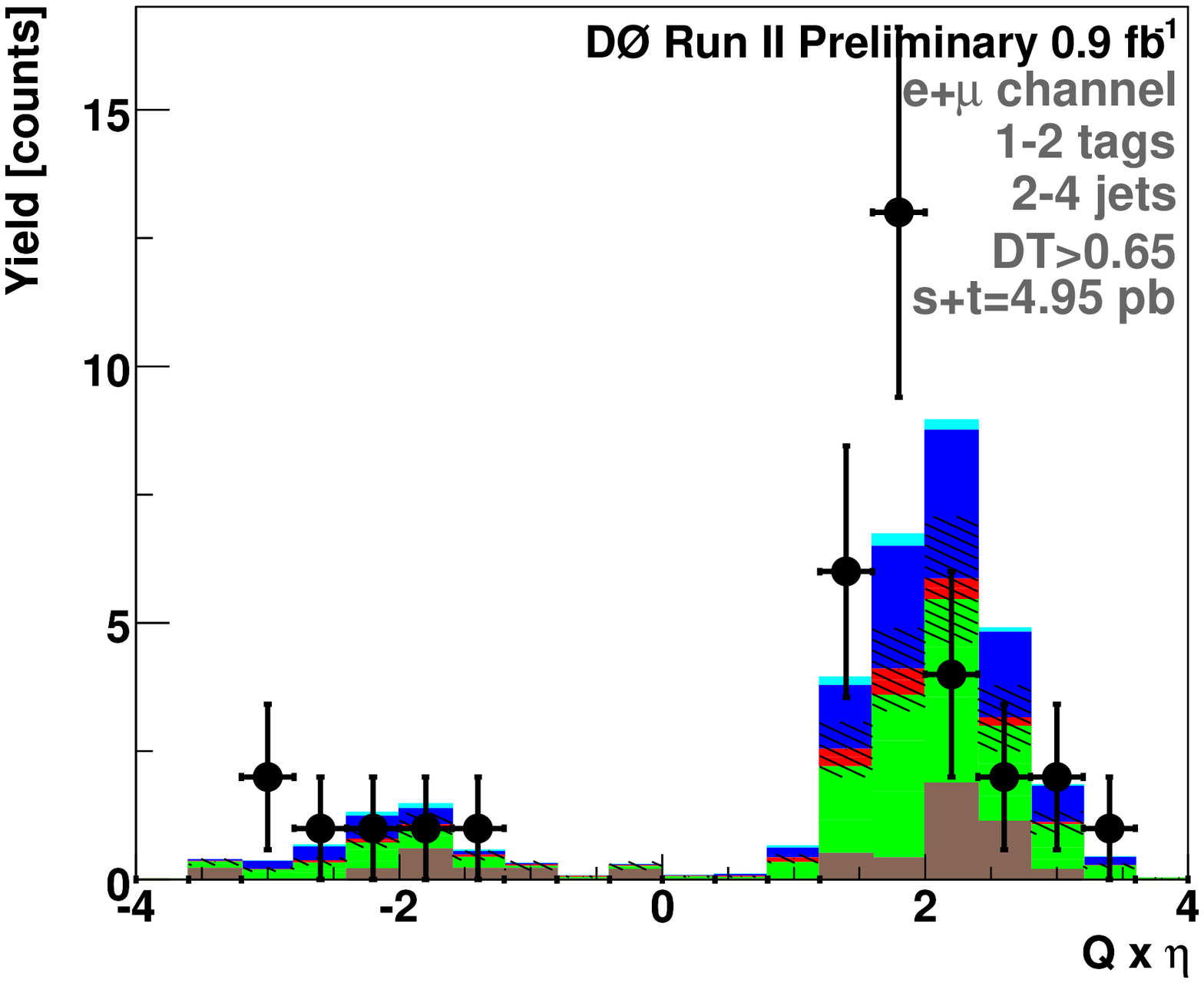}
\vspace{-0.1in}
\caption[dtplots]{Data-background distribution of Q(lepton)$\times\eta$(untagged
leading jet) for three different cuts on the $tb+tqb$ DT output: low (DT$<$0.3,
left plot), high (DT$>$0.55, middle plot), and very high (DT$>$0.65, right plot).}
\label{qeta_plots}
\end{figure*}

To verify the kinematics of the selected events in the most discriminant region
of the DT outputs, the $t$-channel
characteristic distribution of Q(lepton)$\times\eta$(untagged
leading jet) is plotted in different slices of the DT output in
Fig.~\ref{qeta_plots}. Requiring higher DT outputs clearly selects more
signal-like data and the distinct asymmetric signal shape can be seen taking form. 

The three analyses are highly correlated since they all use the same signal and
background models and data, with almost the same systematic uncertainties. The
correlation between the three methods has been measured in fake pseudo data-sets
(which include the systematic uncertainties on our background model), with the
SM single top cross section.  The best linear unbiased estimator 
 (BLUE) has been applied to the three measured values and their correlations
to give a combined measured cross section of $\sigma({\ppbar}{\rargap}tb+X,~tqb+X) = 4.8 \pm
1.3$~pb, which corresponds to a significance of 0.027\% (or 3.5~standard deviations). 
\vspace{-0.4in}

\section{Measurement of $|V_{tb}|$}
We use the decision tree measurement of the $tb$+$tqb$ cross section
to derive a first direct measurement of the strength of the $V-A$
coupling $|V_{tb} f_1^L|$ in the $Wtb$ vertex, where $f_1^L$ is an
arbitrary left-handed form factor~\cite{vtb-theory}. We measure
$|V_{tb} f_1^L| = 1.3 \pm 0.2$. This measurement assumes $|V_{td}|^2 +
|V_{ts}|^2 \ll |V_{tb}|^2$ and a pure $V$$-$$A$ and CP-conserving
$Wtb$ interaction. Assuming in addition that $f_1^L = 1$, we obtain
$0.68 < |V_{tb}| \le 1$ at $95\%$~C.L.. These measurements make no
assumptions about the number of quark families or CKM matrix
unitarity.

\section{Summary}
To summarize, we have performed a search for single top quark
production using 0.9~fb$^{-1}$ of data collected by the \dzero~experiment
at the Tevatron collider. We find an excess of events over the
background prediction in the high discriminant output region from
three analyses and interpret it as evidence for single top quark
production. The excess has a combined significance of 3.5 standard
deviations and the combined first measurement of the single top quark
cross section is: $\sigma({\ppbar}{\rargap}tb+X,~tqb+X) = 4.8 \pm
1.3$~pb.

\section*{Acknowledgments}
I thank the organizers of the Rencontres de Moriond for the 
enjoyable and fruitful atmosphere of the meeting; and the Marie
Curie program of the European Union for funding my accommodation at La Thuile.

\end{document}